\documentclass[twocolumn,twocolappendix]{aastex701}

\shorttitle{{\sc Effortless} I. First Results}
\shortauthors{K. Cao}
\usepackage{CJKutf8}

\usepackage{graphicx}	
\usepackage{amsmath}	
\usepackage{amssymb}	
\usepackage{multirow}

\graphicspath{{figures/}}
\usepackage[OT2,T1]{fontenc}
\DeclareSymbolFont{cyrletters}{OT2}{wncyr}{m}{n}
\DeclareMathSymbol{\Sha}{\mathalpha}{cyrletters}{"58}

\definecolor{F184}{rgb}{0.65,0.1,0.0}
\definecolor{H158}{rgb}{0.35,0.4,0.0}
\definecolor{J129}{rgb}{0.0,0.4,0.35}
\definecolor{Y106}{rgb}{0.0,0.1,0.65}

\begin{document}

\title{Efficient Optimal Image Reconstruction for the Nancy Grace Roman Space Telescope and Beyond: I. First Results with {\sc Effortless}}

\author[orcid=0000-0002-1699-6944]{Kaili Cao (\begin{CJK*}{UTF8}{gbsn}曹开力\end{CJK*})}
\affiliation{Center for Cosmology and AstroParticle Physics (CCAPP), The Ohio State University, 191 West Woodruff Ave, Columbus, OH 43210, USA}
\affiliation{Department of Physics, The Ohio State University, 191 West Woodruff Ave, Columbus, OH 43210, USA}
\email[show]{cao.1191@osu.edu}

\collaboration{all}{Roman HLIS Cosmology PIT}

\begin{abstract}

The forthcoming Nancy Grace Roman Space Telescope will revolutionize astrophysics by generating huge amounts of data of unprecedented quality. To properly address the data deluge and fully realize its potential, analysis tools that are both efficient and optimal are needed. {\sc Effortless} (previously known as Fast {\sc Imcom}) is a new algorithm for linear image reconstruction. Like its predecessor {\sc Imcom}, it offers control over point spread functions in output images; by avoiding laborious calculations, it is tens of times faster and has a smaller memory footprint. In this paper, I apply {\sc Effortless} to simulated Roman images and present promising first results. With ideal point sources, I illustrate that a single image reconstructed by {\sc Effortless}, combined with the post-measurement calibration procedure described in a companion paper, can lead to better measurements than a set of $\sim 6$ images coadded by {\sc Imcom}. While both algorithms were originally designed for weak gravitational lensing cosmology, {\sc Effortless} can benefit studies of static features and dynamic aspects of the Universe alike. Moreover, the efficiency and interpretability of {\sc Effortless} provides new possibilities for further reducing errors in measurements. The implementation of {\sc Effortless} is detailed in the companion paper.

\end{abstract}


\section{Introduction} \label{sec:intro}

NASA's Nancy Grace Roman Space Telescope \citep[hereafter Roman;][]{2019arXiv190205569A, 2025arXiv250510574O} is currently scheduled to launch in August 2026. Along with other facilities like the NSF-DOE Vera C. Rubin Observatory \citep{2012arXiv1211.0310L, 2019ApJ...873..111I} and ESA's Euclid space telescope \citep{2011arXiv1110.3193L, 2022A&A...662A.112E, 2025A&A...697A...1E}, Roman will conduct next-generation astronomical surveys and revolutionize various fields of astrophysics. To fully realize such potential, the community needs to develop efficient analysis tools to address the data deluge while maintaining the needed accuracy and precision.

Roman's High-Latitude Wide-Area Survey (HLWAS) will enhance studies of the large-scale structure of the Universe via weak gravitational lensing \citep[see][for some recent reviews]{2013PhR...530...87W, 2015RPPh...78h6901K, 2018ARA&A..56..393M} with its imaging component (the High-Latitude Imaging Survey, or HLIS in short) and baryon acoustic oscillations \citep[also see][]{2013PhR...530...87W} with its spectroscopic component \citep{2022ApJ...928....1W}. Roman will also advance time-domain sciences with its Galactic Bulge Time-Domain Survey \citep[GBTDS; e.g.,][]{2019ApJS..241....3P, 2023ApJS..269....5W, 2025ApJ...987..181W} and High-Latitude Time-Domain Survey \citep[HLTDS; e.g.,][]{2025ApJ...988...65R, 2025ApJ...993..116K}. Many of these purposes require image reconstruction to improve image quality for studying static objects or facilitate comparisons between different epochs for detecting dynamic events.

Image reconstruction is usually formulated as linear transformations to permit the well-definedness of point spread functions \citep[PSFs;][]{2023OJAp....6E...5M}. Thanks to its robustness and efficiency, the {\sc Drizzle} algorithm \citep{2002PASP..114..144F, 2012drzp.book.....G} has been widely applied to the reconstruction of images taken by space telescopes (which have stable PSFs). However, the Roman weak gravitational lensing cosmology program imposes stringent requirements on the control of systematic errors \citep[e.g.,][]{2026arXiv260100438C}. To achieve PSF uniformity and regularity, my colleagues and I have been developing the {\sc Imcom} (IMage COMbination) algorithm \citep{2011ApJ...741...46R, 2024MNRAS.528.2533H, 2024MNRAS.528.6680Y, 2025ApJS..277...55C, 2026ApJ...998..304C} and validating it with realistic simulated Roman images \citep{2023MNRAS.522.2801T, 2025MNRAS.544.3799O}. Despite its optimality, {\sc Imcom} is orders of magnitude slower than {\sc Drizzle}, limiting its applications in areas other than Roman weak lensing.

In \citet{2026AJ....171..140C}, I introduced a new algorithm, {\sc Effortless} (EFFicient Optimal image ReconsTruction using LESS memory),\footnote{It was previously named Fast {\sc Imcom} after its predecessor, but I realized that it is not necessarily combining images and decided to give it a new name.} and demonstrated its feasibility with 1D experiments. To obtain optimal weights (i.e., coefficients) in linear transformations, {\sc Imcom} builds and solves large linear systems; {\sc Effortless} directly yields solutions based on understanding of those matrices, thus it is much more efficient and hopefully at least as optimal. While the 2D implementation will be detailed in my companion paper (K. Cao 2026, in preparation), this paper exhibits some first practical results with the {\sc Effortless} algorithm. The paper is structured as follows. In Section~\ref{sec:meth}, I briefly review the mathematical formalism of {\sc Effortless} and relevant parts of the {\sc Imcom} ecosystem. Then in Section~\ref{sec:res}, I present example reconstructed images and preliminary analysis results. Conclusions, implications, and topics for future work are discussed in Section~\ref{sec:disc}.

\section{Methods} \label{sec:meth}

This section is a brief recap of the {\sc Effortless} formalism (Section~\ref{ss:meth_el}) and the {\sc Imcom} methodology (Section~\ref{ss:meth_ic}). For an in-depth discussion of the underlying mathematical concepts, see Sections~2 and 3 of \citet{2026AJ....171..140C}; for a more complete overview of {\sc PyImcom}, the current implementation of {\sc Imcom}, see Section~2 of \citet{2026ApJ...998..304C}.

\subsection{{\sc Effortless} Formalism} \label{ss:meth_el}

A linear image reconstruction scheme is formulated as a linear transformation from input signals $I_{\boldsymbol i}^{({\bar i})}$ to each output signal $H_{\boldsymbol \alpha}$ (the pixel indices ${\boldsymbol i}$ and ${\boldsymbol \alpha}$ are $2$-tuples of integers since astronomical images are 2D):
\begin{equation}
    H_{\boldsymbol \alpha} = \sum_{{\bar i}} {\cal N}_{{\bar i}} \sum_{{\boldsymbol i} \in {\bar i}} T_{{\boldsymbol \alpha} {\boldsymbol i}}^{({\bar i})} I_{\boldsymbol i}^{({\bar i})},
    \label{eq:transform}
\end{equation}
where ${\bar i} \subset {\mathbb Z}^2$ is the collection of available pixel indices in an input image, ${\cal N}_{{\bar i}}$ is the meta-weight assigned to input image ${\bar i}$, and $T_{{\boldsymbol \alpha} {\boldsymbol i}}^{({\bar i})}$ are the reconstruction weights.\footnote{In \citet{2026AJ....171..140C}, I had $T_{{\boldsymbol \alpha} {\boldsymbol i}}^{({\bar i})} \equiv {\cal N}_{{\bar i}} T_{{\boldsymbol \alpha} {\boldsymbol i}}^{\prime ({\bar i})}$, where $T_{{\boldsymbol \alpha} {\boldsymbol i}}^{\prime ({\bar i})}$ is the same as $T_{{\boldsymbol \alpha} {\boldsymbol i}}^{({\bar i})}$ here (the prime $'$ is not used for this purpose in this paper). Both notations are self-consistent.} The summation over input images does not apply to the scenario where one reconstructs an individual input image.

By assigning weights to input pixels, such a linear transformation also constructs an output PSF $\Psi'_{\boldsymbol \alpha}$ from properly shifted input PSFs $G_{\boldsymbol i}^{\prime ({\bar i})}$ (both are functions of the relative position ${\boldsymbol s}$):\footnote{The prime $'$ in $\Psi'_{\boldsymbol \alpha}$, $G_{\boldsymbol i}^{\prime ({\bar i})}$, and $\Gamma'$ (see below) stands for a ``backward'' PSF as defined in \citet{2026AJ....171..140C}. While the concept of a ``forward'' PSF is not directly used in this paper, the prime $'$ is retained to avoid confusion. In Equation~(7) of \citet{2026AJ....171..140C}, the signs of ${\boldsymbol r}_{\boldsymbol i}$ (the superscript $({\bar i})$ was missing) and ${\boldsymbol R}_{\boldsymbol \alpha}$ needed to be reversed for the input PSFs to be ``properly shifted.'' I apologize for not having realized this before the publication of that paper; the numerical results therein are still valid as the input PSFs were symmetric. I will systematically explain these concepts in K. Cao (2026, in preparation).}
\begin{equation}
    \Psi'_{\boldsymbol \alpha} ({\boldsymbol s}) = \sum_{{\bar i}} {\cal N}_{{\bar i}} \sum_{{\boldsymbol i} \in {\bar i}} T_{{\boldsymbol \alpha} {\boldsymbol i}}^{({\bar i})} G_{\boldsymbol i}^{\prime ({\bar i})} ({\boldsymbol R}_{\boldsymbol \alpha} - {\boldsymbol r}_{\boldsymbol i}^{({\bar i})} + {\boldsymbol s}),
    \label{eq:recons_pixel}
\end{equation}
where ${\boldsymbol R}_{\boldsymbol \alpha}$ and ${\boldsymbol r}_{\boldsymbol i}^{({\bar i})}$ are the positions of output pixel ${\boldsymbol \alpha}$ and input pixels ${\boldsymbol i}$ in input image ${\bar i}$, respectively. It is important to note that {\sc Imcom} and {\sc Effortless} assume perfect modeling of input PSFs, including chromatic effects and spatial or temporal variation. Studying how mismodeling impacts results is left for future work; see Section~6.1 of \citet{2026AJ....171..140C} for some discussion.

To a reasonable approximation, in the vicinity of a given position, $G'$ is the same for all pixels in each input image, i.e., $G_{\boldsymbol i}^{\prime ({\bar i})}$ can be replaced by $G^{\prime ({\bar i})}$. Under this approximation, for each input image, {\sc Effortless} computes reconstruction weights for input pixels by sampling the weight field $T^{({\bar i})}$:
\begin{equation}
    T_{{\boldsymbol \alpha} {\boldsymbol i}}^{({\bar i})} = T^{({\bar i})} ({\boldsymbol r}_{\boldsymbol i}^{({\bar i})} - {\boldsymbol R}_{\boldsymbol \alpha}), \quad {\tilde T}^{({\bar i})} = {\tilde \Gamma}' / {\tilde G}^{\prime ({\bar i})},
    \label{eq:T_solution}
\end{equation}
where ${\tilde \cdot}$ denotes Fourier transform, and $\Gamma'$ is the user-specified target output PSF, which is typically chosen as a Gaussian that is slightly wider than input PSFs in each band \citep{2026ApJ...998..304C}. The resulting output PSFs $\Psi'_{\boldsymbol \alpha}$ are close to the target output PSF $\Gamma'$, and the discrepancies due to finite sampling follow simple patterns \citep[see Figure~5 of][]{2026AJ....171..140C}. While combining input images, there are different strategies to determine the meta-weights ${\cal N}_{{\bar i}}$; here I adopt the $\Sigma$-first strategy defined and advocated in \citet{2026AJ....171..140C}: assigning equal meta-weights to all input images available to each output pixel.

Implementation of this algorithm, including practical issues like geometric distortions, numerical artifacts, input pixel masks, etc., will be detailed in my companion paper (K. Cao 2026, in preparation).

\subsection{{\sc Imcom} Methodology} \label{ss:meth_ic}

For linear image reconstruction algorithms like {\sc Imcom} and {\sc Effortless}, the reconstruction weights $T_{{\boldsymbol \alpha} {\boldsymbol i}}^{({\bar i})}$ and the meta-weights ${\cal N}_{{\bar i}}$ (if applicable) do not depend on the input signals $I_{\boldsymbol i}^{({\bar i})}$. Hence the same weights can be used to reconstruct different versions of input images generated for different purposes; these are called input ``layers'' in the {\sc Imcom} terminology \citep[see Section~3 of][for further details]{2024MNRAS.528.2533H}.

To facilitate comparisons, for simulations in this paper, I use the same inputs as those in \citet{2026ApJ...998..304C}\footnote{The only exception is that there is a minor difference concerning input pixel masks. In \citet{2026ApJ...998..304C}, in addition to input pixels that are known to be inoperable or are affected by simulated cosmic rays, my colleagues and I masked another $\sim 0.1\%$ of the input pixels that have high noise levels according to laboratory tests of Roman detectors. However, the version of laboratory noise data that is ready for use by {\sc PyImcom} (and thus {\sc Effortless}) was not available as of simulations in this work. Therefore, I slightly increased the simulated cosmic ray rate (by a factor of $1.2$) so that the numbers of masked input pixels are roughly matched to those in \citet{2026ApJ...998..304C}. This ensures that results in this paper do not have an undue advantage when compared to those in that paper.} and the same target outputs as the benchmark cases of the Cholesky linear algebra kernel therein. Specifically, four input layers are used: simulated science images \citep[{\tt \textquotesingle SCI\textquotesingle};][]{2025MNRAS.544.3799O}, injected stars drawn by {\sc GalSim} \citep[{\tt \textquotesingle gsstar14\textquotesingle};][]{2015A&C....10..121R}, and simulated white ({\tt \textquotesingle whitenoise10\textquotesingle}) and $1/f$ ({\tt \textquotesingle 1fnoise9\textquotesingle}) noise frames \citep{2024MNRAS.528.2533H}. The output configuration is summarized in Tables~1 and 4 of \citet{2026ApJ...998..304C}; due to differences between {\sc Imcom} and {\sc Effortless}, some of the {\sc Imcom} hyperparameters do not apply, and a few {\sc Effortless} hyperparameters need to be added. Such differences do not affect the physical meaning of output images and will be explained in K. Cao (2026, in preparation). In this paper, I use {\sc Effortless} to reprocess the $16$ blocks of size $(1.75 \,{\rm arcmin})^2$ as shown in Figure~1 of \citet{2026ApJ...998..304C} in the Y106, J129, H158, and F184 bands.\footnote{In the current design of the Roman HLIS \citep{2025arXiv250510574O}, the Medium Tier ($2415 \,{\rm deg}^2$) will use the Y106, J129, and H158 filters, and the Wide Tier ($2702 \,{\rm deg}^2$) will only use the H158 band. Here the F184 band is included for consistency, and the K213 band \citep[which was in][]{2026ApJ...998..304C} is not included.}

The preliminary analysis in this work focuses on moment-based measurements of the injected stars (simulated ideal point sources). The measurements are conducted with the {\sc HSM} \citep{2003MNRAS.343..459H, 2005MNRAS.361.1287M} module of {\sc GalSim} and include amplitude $A$ (zeroth moment), centroid offset ${\boldsymbol d}$ (first moments), shear invariant width $s$ and ellipticity $g_1$ and $g_2$ (based on second moments), and the spin-$2$ fourth moment $M^{\rm (4)}_{\rm PSF}$ \citep{2023MNRAS.520.2328Z, 2023MNRAS.525.2441Z}. See Section~2.3 of \citet{2026ApJ...998..304C} for their respective definitions. Most of these quantities are expected to be $0$ for injected stars, and expected values of $A$ and $s$ are determined by the chosen target output PSFs. Raw measurements on individual images reconstructed by {\sc Effortless} have relatively large errors due to undersampling, but leveraging the simple patterns manifested by these errors, they can be largely calibrated out. Such post-measurement calibration is another major topic of K. Cao (2026, in preparation). In short, the calibration amounts to detrending trigonometric functions of subpixel positions like $\cos (2\pi \Delta x)$ using a linear model; see Section~6 of that paper for details.

\section{Results} \label{sec:res}

Since {\sc Effortless} has ``efficient'' and ``less memory'' in its name, it is natural to discuss consumption of computational resources. Since the first square degree-scale simulations with {\sc PyImcom} in the summer of 2024 \citep{2025ApJS..277...55C, 2026ApJ...998..304C}, further software improvements and hyperparameter optimization have made it even more efficient. Besides, the CCAPP condo\footnote{\url{https://www.osc.edu/supercomputing/computing/cardinal/ccapp_condo}} at the Ohio Supercomputer Center \citep{OhioSupercomputerCenter1987} has been migrated from the Pitzer cluster \citep{Ohio_Supercomputer_Center2018-dl} to the more powerful Cardinal cluster \citep{Ohio_Supercomputer_Center2024-dl}. Combining these factors, the current quote for {\sc PyImcom} performance is $\sim 1.1 \times 10^4$ core hours per filter per square degree for the Medium and Wide Tiers of Roman HLIS (with a typical coverage of $\sim 6$ per band). While introducing {\sc Effortless} in \citet{2026AJ....171..140C}, my ``educated speculation'' was that {\sc Effortless} would be ``an order of magnitude faster than {\sc Imcom}.'' It turns out that such speculation is rather conservative, and it only takes {\sc Effortless} $9$--$13$ core minutes to process a block of size $(1.75 \,{\rm arcmin})^2$, i.e., $\sim 2 \times 10^2$ core hours per filter per square degree, or $50$--$60$ times faster than {\sc PyImcom}. Meanwhile, {\sc Effortless} has a small memory footprint, typically below $\sim 2 \,{\rm GiB}$ for simulations in this work, while {\sc PyImcom} may exceed $\sim 4 \,{\rm GiB}$ to process the same data. In the rest of this section, I present example reconstructed images and preliminary analysis results to demonstrate that less time and memory does not necessarily mean lower quality.

\begin{figure*}
    \centering
    \includegraphics[width=\textwidth]{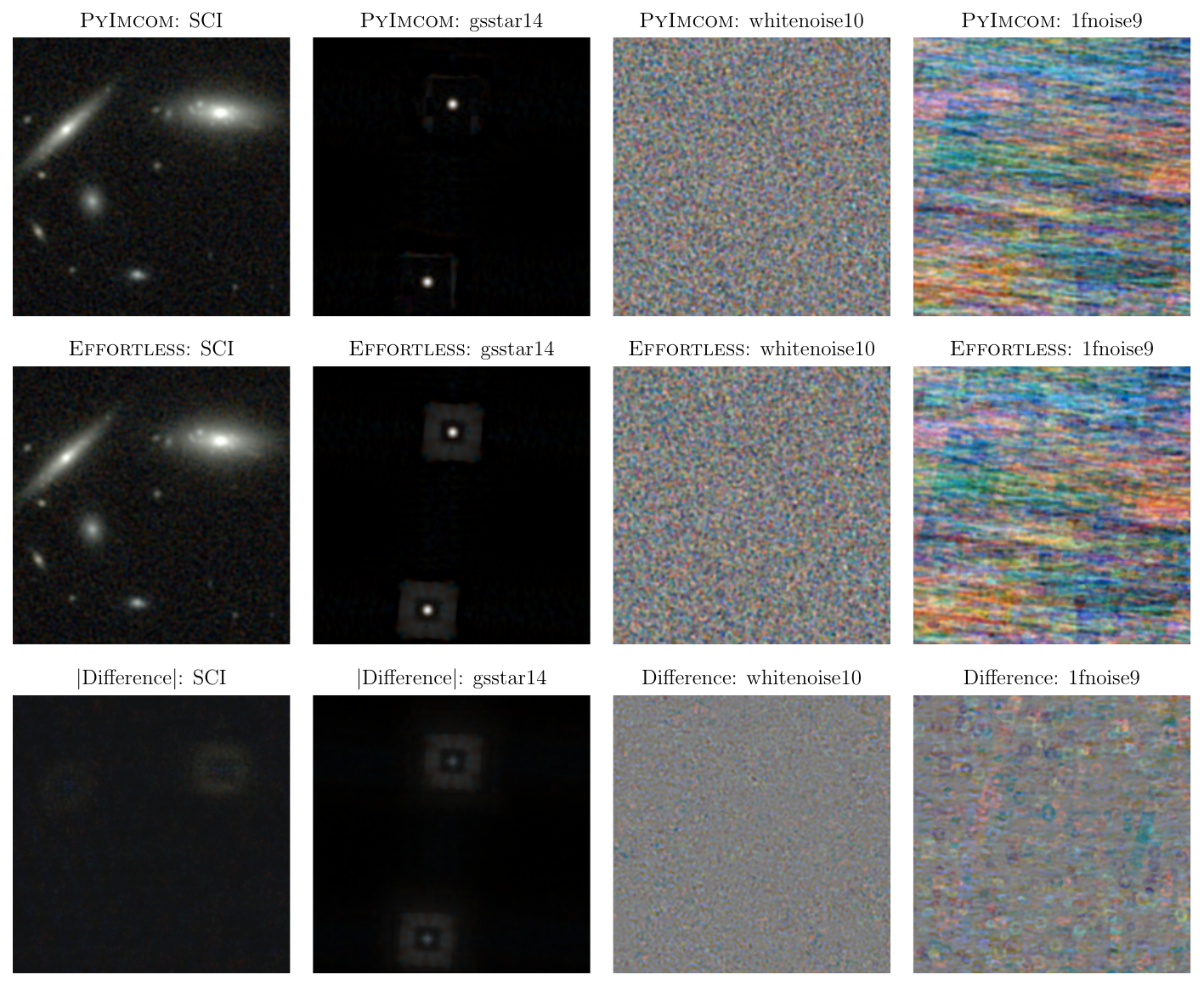}
    \caption{Four layers in a field of $17.5 \,{\rm arcsec}$ ($448$ output pixels) on a side, coadded by the Cholesky kernel of {\sc PyImcom} (upper row) and the $\Sigma$-first strategy of {\sc Effortless} (middle row), along with the differences between them (absolute difference for the first two columns, {\sc PyImcom} minus {\sc Effortless} for the last two columns; lower row). Images in the upper row are identical to those in the upper row of Figure~3 of \citet{2026ApJ...998..304C}. Each panel is a Y106 ({\color{Y106} \#001AA6}) + J129 ({\color{J129} \#006659}) + H158 ({\color{H158} \#596600}) + F184 ({\color{F184} \#A61A00}) composite; note that these four colors have similar lightnesses and add up to white (\#FFFFFF). From left to right, the four layers are: simulated science images ({\tt \textquotesingle SCI\textquotesingle}), injected stars drawn by {\sc GalSim} ({\tt \textquotesingle gsstar14\textquotesingle}), simulated white noise frames ({\tt \textquotesingle whitenoise10\textquotesingle}), and simulated $1/f$ noise frames ({\tt \textquotesingle 1fnoise9\textquotesingle}). The scaling is set following Figure~8 of \citet{2024MNRAS.528.2533H} for {\tt \textquotesingle SCI\textquotesingle} and following Figure~1 of \citet{2024MNRAS.528.6680Y} for the other three layers.
    \label{fig:images_all}}
\end{figure*}

Figure~\ref{fig:images_all} shows example four-band composites of {\sc PyImcom} (upper row) and {\sc Effortless} (middle row). The purpose of this figure is to qualitatively illustrate that the two algorithms produce similar results; quantitative comparisons to the ``truth'' will be conducted later in this section. Besides, the recommended way of using {\sc Effortless} is to reconstruct individual images instead of coadding them, as coaddition entangles defects from different input images and makes them difficult to track and reduce.

In most cases, the differences between them (lower row) are not very obvious via visual inspection. Both algorithms are able to reconstruct simulated science images (first column), and it is expected that source detection algorithms that can work with one of them can work with the other. Injected stars (second column) have been used as a ``stress test'' for {\sc Imcom}. Both algorithms are able to reconstruct their central parts with high fidelity; a closer look would reveal faint residuals in the difference panel (exaggerated by the chosen scaling; see below for further information). Like in the iterative kernel results in \citet{2025ApJS..277...55C, 2026ApJ...998..304C}, ``PSF patches'' (i.e., how ``stars'' are ``injected''; see below for more) can be seen in {\sc Effortless} outputs, but outer regions are arguably less important for moment-based measurements using weight functions vanishing at large distances. The reconstructed versions of input white noise (third column) are again very similar to each other; as for input $1/f$ noise (last column), the two algorithms have slightly different responses to regions surrounding bad input pixels (seen as little circles in the difference panel). In such regions, {\sc PyImcom} tries to mitigate the impact of bad input pixels, while {\sc Effortless} simply excludes the corresponding input image(s). Further investigations of noise properties of reconstructed images and their impact on measurements are left for future work \citep[but see][for some previous work]{2024MNRAS.528.6680Y, 2024PASP..136l4506L}.

\begin{figure*}
    \centering
    \includegraphics[width=\textwidth]{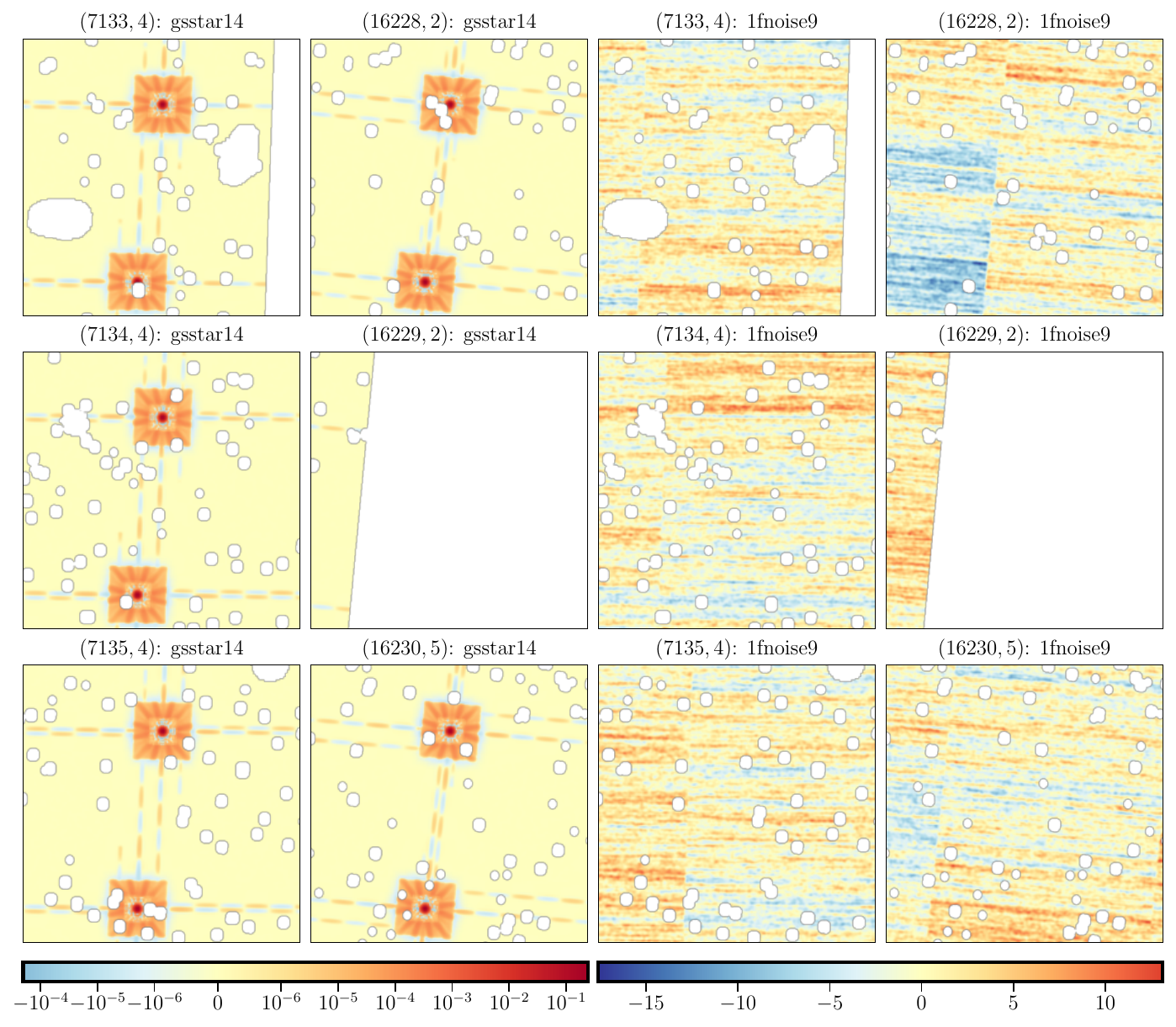}
    \caption{Two layers in the same field as Figure~\ref{fig:images_all}, reconstructed by {\sc Effortless} from six individual images in the Y106 band. The left half shows injected stars drawn by {\sc GalSim} ({\tt \textquotesingle gsstar14\textquotesingle}) in symmetric logarithmic scale with a linear threshold of $10^{-6}$; the right half shows simulated $1/f$ noise frames ({\tt \textquotesingle 1fnoise9\textquotesingle}) in linear scale. Masked output pixels are shown in white.
    \label{fig:images_Y106}}
\end{figure*}

Figure~\ref{fig:images_Y106} presents two layers of individual images reconstructed by {\sc Effortless}. These images are in the Y106 band, which is the bluest band for most of the Roman weak lensing program and thus is subject to the most severe undersampling effects (smallest wavelength-to-aperture diameter ratio $\lambda/D$ for the same native pixel scale). It is worth emphasizing that {\sc Effortless} can project individual images onto a common pixel grid and unify the PSFs; such capability (which {\sc Imcom} is not thought to have\footnote{According to the Nyquist--Shannon sampling theorem, a linear algorithm to reconstruct an oversampled image with a uniform PSF from a single undersampled native image does not exist, and multiple undersampled images are needed for that purpose \citep[see Appendix~C of][]{2024MNRAS.528.2533H}. However, what {\sc Effortless} accomplishes is slightly different from what we said was impossible: {\sc Effortless} reconstructs an output image from a single input image with a PSF whose shape has ``rippling'' (depending on subpixel positions of sources) that can be calibrated. Since {\sc Effortless} has been demonstrated to have such capability, testing whether {\sc Imcom} also does seems moot.}) enables pixel-by-pixel comparisons between different images for obtaining time-domain information or removing some exposure-specific errors, although the efficacy of such comparisons can be limited by PSF residuals due to imperfections of the image reconstruction process. From the distributions of masked output pixels (shown in white), one can easily see the positions of bad input pixels and the boundaries of Roman detectors; but outside the corresponding regions, the {\sc Effortless} reconstruction is robust against the unavailability of some input pixels.

The left half of Figure~\ref{fig:images_Y106} shows injected stars in symmetric logarithmic scale. In such a scale, one can clearly see square edges and diffraction spikes in the outer regions of injected stars, as well as alternating positive and negative features in the central parts. {\sc PyImcom} injects these point sources by drawing (resampled) PSF images onto that layer. Due to the relative small size of the PSF images ($32^2$ native pixels in \citet{2026ApJ...998..304C} and this work), the square edges are prominent. These edges are not a physical feature of Roman PSFs and are already present in the {\tt \textquotesingle gsstar14\textquotesingle} before the reconstruction process. Outer residuals, especially diffraction spikes, of bright objects (usually stars in the Milky Way) are worth removing so that their faint neighbors can be measured; this can be done via either simply increasing the window size for selecting input pixels or subtracting properly scaled PSFs from input images (E. Macbeth et al. 2026, in preparation). Since the former leads to higher computational costs and may not be robust against geometric distortions, the latter is probably preferred; {\sc Effortless} can simplify and accelerate the workflow and facilitate the characterization of bright objects.\footnote{The alternating positive and negative stripes outside PSF patches, previously seen in Figure~16 of \citet{2025ApJS..277...55C}, are artifacts in the input data, as {\sc Imcom} and {\sc Effortless} only assign non-zero weights to input pixels in the vicinity of each output pixel. Such artifacts are only seen in this kind of logarithmic plots and otherwise would have negligible impact in the presence of noise or background.} The central alternating features are artifacts of the reconstruction process; fortunately, they follow simple patterns \citep[see Figure~5 of][]{2026AJ....171..140C}, and their impact on measurements of injected point sources can be largely calibrated out (which is why {\sc Effortless} can outperform {\sc Imcom} in terms of accuracy, see below). 

The right half of Figure~\ref{fig:images_Y106} shows simulated $1/f$ noise frames as another example of the use cases of pixel-by-pixel comparisons mentioned above. Since the structure of correlated noise is retained, removal of correlated noise stripes \citep[e.g.,][]{2026PASJ...78..810L} can be done by comparing reconstructed images at the square arcminute level (i.e., expensive joint processing at the square degree level can be avoided). Exploring these new possibilities opened up by {\sc Effortless} is left for future work.

\begin{figure*}
    \centering
    \includegraphics[width=\textwidth]{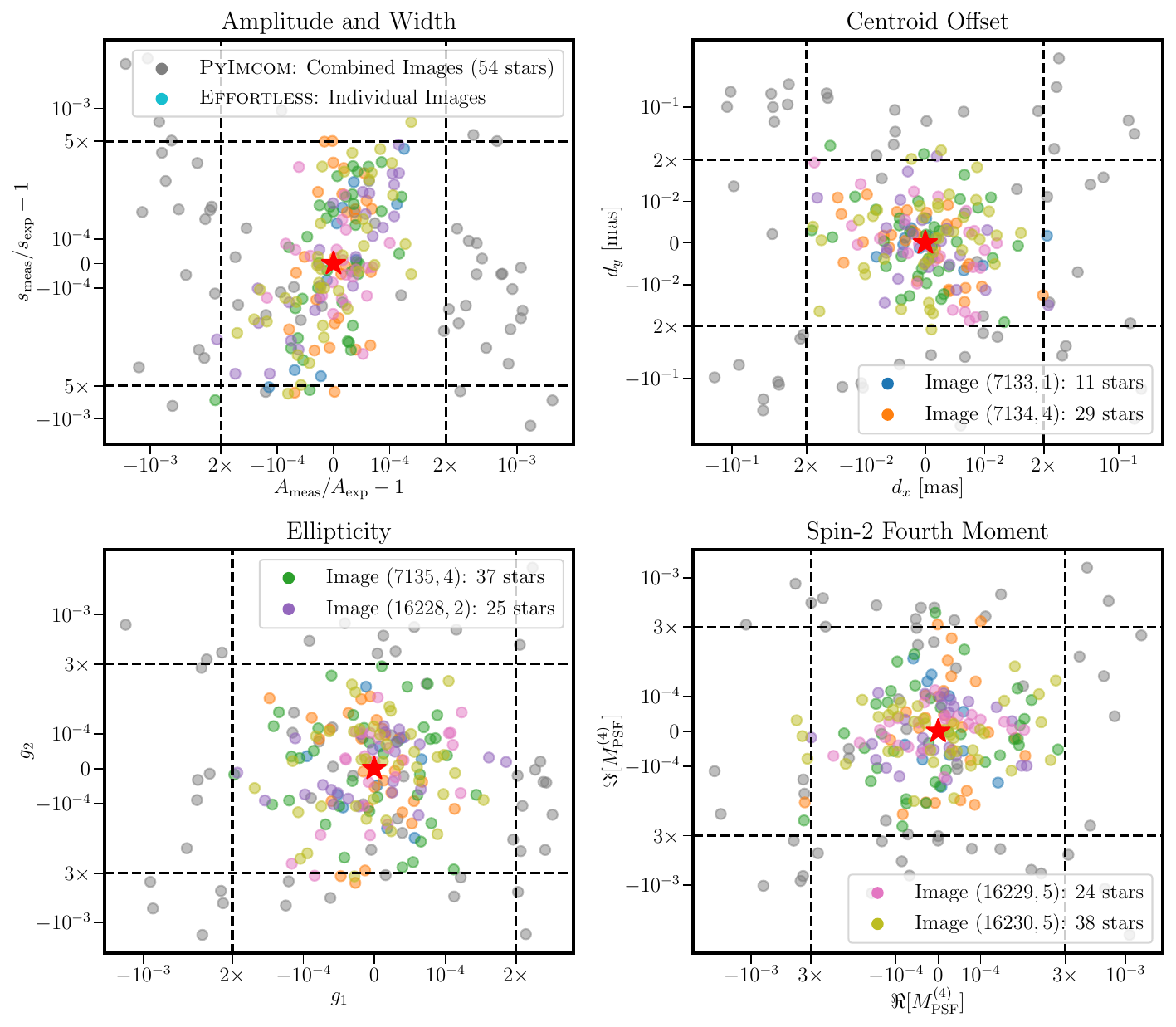}
    \caption{{\sc HSM} measurements of the $54$ injected stars in block $(14, 12)$ in the Y106 band. {\sc PyImcom} results based on the combination of $\sim 6$ images are shown in gray; calibrated {\sc Effortless} results based on the reconstruction of $6$ individual images are shown in different colors. Numbers of stars measured from individual images are shown in the legends; measurements of each injected star are provided by at least $1$ image. The four panels correspond to amplitude $A$ and shear invariant width $s$ (upper left panel), centroid offset ${\boldsymbol d}$ (upper right panel), ellipticity $g_1$ and $g_2$ (lower left panel), and the spin-$2$ fourth moment $M^{\rm (4)}_{\rm PSF}$ (lower right panel), respectively. For $A$ and $s$, fractional errors (measured divided by true minus $1$) are shown. All axes are in symmetric logarithmic scales, with different linear thresholds (shown as vertical and horizontal black dashed lines) chosen to enclose most of {\sc Effortless} data points. The axis tick label ``$n \times$'' means ``$n \times$'' the value of the closest inner label, e.g., $\pm 2 \times 10^{-4}$ for the $x$-axis of the upper left panel. Correct values are denoted as red stars.
    \label{fig:analysis_Y106}}
\end{figure*}

Figure~\ref{fig:analysis_Y106} shows {\sc HSM} measurements of $54$ injected stars in a block (which includes the field shown in Figures~\ref{fig:images_all} and \ref{fig:images_Y106}) in the Y106 band. Again, the Y106 band is chosen because it is the most challenging; the block and input images are not special in any way. {\sc Effortless} results have been calibrated leveraging patterns of PSF residuals (see Section~\ref{ss:meth_ic}). Such calibration can be viewed as a counterpart to the combination of input images in {\sc Imcom}, as both are ways of mitigating undersampling effects. In principle, it is possible to calibrated {\sc Imcom} results using subpixel positions in input images as well; nonetheless, since the matrix inversion process in {\sc Imcom} is largely a black box, the simple patterns followed by PSF residuals are likely corrupted, and how to perform such a calibration is unclear.

Using individual images reconstructed by {\sc Effortless} (combined with the calibration), absolute values of the measurement errors are roughly bounded by $2 \times 10^{-4}$ (fractional) for $A$, $5 \times 10^{-4}$ (fractional) for $s$, $2 \times 10^{-2} \,{\rm mas}$ for both components of centroid offset ${\boldsymbol d}$, $2 \times 10^{-4}$ and $3 \times 10^{-4}$ for the two components of ellipticity $g_1$ and $g_2$, and $3 \times 10^{-4}$ for both parts of the spin-$2$ fourth moment $M^{\rm (4)}_{\rm PSF}$. In light of input pixel masks, all these stars cannot be measured via each (reconstructed) input image. On average, each of the $54$ stars has $3.04$ sets of measurements from individual images reconstructed by {\sc Effortless}; $4$ of them only have $1$, and $3$ of them have $5$. When multiple sets of measurements are available, it is expected that combining them can lead to even better results.\footnote{Since measurement errors from different input images may be correlated, they cannot be simply averaged out. Devising a concrete way of combining measurements is beyond the scope of this paper.}

Shown in gray are measurements of the same stars based on the combined image produced by {\sc PyImcom} \citep[taken from][]{2026ApJ...998..304C}. While the errors are within the requirements for the Roman weak gravitational lensing cosmology program \citep{2024MNRAS.528.6680Y},\footnote{My colleagues and I have been assuming that point sources constitute a ``stress test'' for image reconstruction algorithms. Testing whether measurements of extended sources also meet those requirements is work in progress.} they are up to an order of magnitude larger than those based on {\sc Effortless} outputs. Since the high quality of {\sc Effortless} results relies on a post-measurement calibration based on the subpixel positions of injected stars in different input images, I tentatively conclude that it may make more sense to conduct measurements on individual reconstructed images (where PSF residuals are better understood) and combine the results, rather than to conduct measurements on combined images. A detailed description of the calibration process and a more comprehensive view of the analysis results will be presented in K. Cao (2026, in preparation).

Before concluding this section, it is worth emphasizing that the above results are based on noiseless point sources drawn using the same input PSFs that are used for image reconstruction. Future work is needed to enable precise and accurate measurements of noisy extended sources (like galaxies in real-world observations) without PSFs known a priori. Nevertheless, \citet{2026AJ....171..140C} and this paper lay the foundation for efficient and optimal image analysis using {\sc Effortless}.

\section{Discussion} \label{sec:disc}

In this paper, I have applied {\sc Effortless} \citep[EFFicient Optimal image ReconsTruction using LESS memory, previously known as Fast {\sc Imcom};][]{2026AJ....171..140C}, a new algorithm for linear image reconstruction, to realistic simulated Roman images \citep{2025MNRAS.544.3799O}. With the same inputs and target outputs as a suite of previous tests of {\sc PyImcom} \citep{2011ApJ...741...46R, 2024MNRAS.528.2533H, 2024MNRAS.528.6680Y, 2025ApJS..277...55C, 2026ApJ...998..304C}, I have found that {\sc Effortless} is $50$--$60$ times faster for the Medium and Wide Tiers of Roman HLIS \citep{2025arXiv250510574O}. With ideal point sources measured by the {\sc HSM} \citep{2003MNRAS.343..459H, 2005MNRAS.361.1287M} module of {\sc GalSim} \citep{2015A&C....10..121R}, I have illustrated that a single image reconstructed by {\sc Effortless}, combined with a simple post-measurement calibration procedure, can lead to better measurements than a set of $\sim 6$ images coadded by {\sc Imcom}.

In addition to being a more efficient and accurate successor to {\sc Imcom}, {\sc Effortless} opens up a myriad of new possibilities. For example, the capability of projecting individual images onto a common pixel grid and unifying the PSFs enables pixel-by-pixel comparisons between different epochs. Other possibilities include but are not limited to semi-analytic models for understanding properties of reconstructed images, simple calculations for optimization of the tiling pattern, and iterative corrections to astrometry, flux calibration, and PSF modeling. See Section~6.2 of \citet{2026AJ....171..140C} for a fuller discussion about potential applications of {\sc Effortless}. Meanwhile, to enable reliable measurement of noisy extended sources (instead of ideal point sources), some challenges remain to be overcome (see Section~6 of K. Cao 2026, in preparation). These are important directions for future work regarding {\sc Effortless}.

\section*{Acknowledgments}

I appreciate useful discussions with colleagues in the Shear and Clustering Measurement (SCM) Working Group of the Roman High Latitude Imaging Survey (HLIS) Cosmology Project Infrastructure Team (PIT) and the Roman Alerts Promptly from Image Differencing (RAPID) Project Infrastructure Team (PIT). This paper has undergone internal review in the Roman HLIS Cosmology PIT. I would like to thank Christopher M. Hirata, Axel Guinot, M.~A.~Troxel, and Mike Jarvis for helpful comments and feedback during the review process.  

This work was supported by the ``Maximizing Cosmological Science with the Roman High Latitude Imaging Survey'' Roman Project Infrastructure Team (NASA grant 22-ROMAN11-0011). I thank Christopher M. Hirata for making the OpenUniverse2024 data needed for this project available on the Ohio Supercomputer Center. The detector mask files used in the Roman image simulations are based on data acquired in the Detector Characterization Laboratory (DCL) at the NASA Goddard Space Flight Center. My colleagues and I would like to thank the personnel at the DCL for making the data available for this project.

Computations for this project used the Cardinal cluster at the Ohio Supercomputer Center \citep{Ohio_Supercomputer_Center2024-dl}. This project made use of the {\sc NumPy} \citep{2020Natur.585..357H}, {\sc Astropy} \citep{2013A&A...558A..33A, 2018AJ....156..123A, 2022ApJ...935..167A}, {\sc Numba} \citep{2015llvm.confE...1L}, {\sc SciPy} \citep{2020NatMe..17..261V}, and {\sc scikit-learn} \citep{2011JMLR...12.2825P} packages. Some of the results in this paper were derived using the {\sc healpy} package \citep{2019JOSS....4.1298Z} based on the {\sc HEALPix} scheme \citep{2005ApJ...622..759G}. Figures in this paper were made using {\sc Matplotlib} \citep{Hunter:2007}; {\sc SAOImageDS9} \citep{2003ASPC..295..489J} played an important role as a preview tool.

\section*{Data Availability}

The OpenUniverse2024 data are publicly available in the NASA/IPAC Infrared Science Archive.\footnote{\url{https://irsa.ipac.caltech.edu/data/theory/openuniverse2024/overview.html}} The {\sc Effortless} (previously known as Fast {\sc Imcom}) and {\sc PyImcom} software packages are available in the following GitHub repositories:
\begin{itemize}
    \item \url{https://github.com/Roman-HLIS-Cosmology-PIT/effortless.git}
    \item \url{https://github.com/Roman-HLIS-Cosmology-PIT/pyimcom.git}
\end{itemize}

Specifically, this work used {\sc Effortless} v0.2.2 and {\sc PyImcom} v1.0.3. The frozen versions are available on Zenodo under an open-source Creative Commons Attribution license: \dataset[doi: 10.5281/zenodo.20365266]{https://doi.org/10.5281/zenodo.20365266} and \dataset[doi: 10.5281/zenodo.17832923]{https://doi.org/10.5281/zenodo.17832923}.

\bibliography{main}{}
\bibliographystyle{aasjournalv7}

\end{document}